\documentclass[12pt]{iopart}

\usepackage{graphicx}
\usepackage{subcaption}
\usepackage{stackengine}
\usepackage{microtype}
\usepackage[utf8]{inputenc}
\usepackage[breaklinks=true]{hyperref}

\usepackage{xcolor}
\hypersetup{
    colorlinks,
    linkcolor={red!50!black},
    citecolor={blue!50!black},
    urlcolor={blue!80!black}
}

\begin{document}

\title{Velocity of interfaces with short and long ranged elasticity under sinusoidal creep}

\author{Juha Savolainen$^1$ and Mikko Alava$^{1,2}$}
\address{$^1$Aalto University, Department of Applied Physics, PO Box 11000, 00076 Aalto, Finland}
\address{$^2$NOMATEN Centre of Excellence, National Centre for Nuclear Research, A. Soltana 7, 05-400 Otwock-Swierk, Poland}

\ead{juha.2.savolainen@aalto.fi}
\vspace{10pt}
\begin{indented}
\item[]\today
\end{indented}

\begin{abstract}
Plenty of research on elastic interfaces has been done on systems where the interface is pushed with a constant force. We studied the average velocity of an interface under a sinusoidal driving in the creep region, considering both short-range elastic systems, such as magnetic domain walls during a hysteresis loop, and long-ranged systems such as fractures. We obtained a modified version of the creep velocity with approximate power-law behaviour and a material-dependent exponent for short ranged systems and simpler behaviour for long-range elasticity. We discuss whether the model can be applied to fatigue fractures or if extra physics is needed.
\end{abstract}

%
\vspace{2pc}

\section{Introduction}

For a large enough ferromagnet, the magnetization is not uniform, but rather the object divides into regions called magnetic domains. The spins of particles in each domain align roughly in the same direction, but neighboring domains will have different alignments, resulting in competing sub-magnets that together create the magnetization of the whole object.

Although the domains are affected by the structure of the magnet, they are not static but rather change with an external magnetic field and thermal noise. Between the domains are thin strips called domain walls, where the spins point somewhere in between the directions of the spins inside the domains. The spins at these boundaries are easier to change than the whole domains, so the magnetization of a multi-domain magnet is changed by moving the domain walls, increasing the size of the domains that align with the applied external magnetic field. \cite{weiss1906, LandaLifshitz8} The magnetization of a small single-domain magnet is harder to change, as the magnetization of the whole object has to be changed at once \cite{StonerWolfarth}.

Domain walls are an example of elastic interfaces. Elastic interfaces describe a variety of phenomena, ranging from wetting lines separating dry and wet regions in a material \cite{Fluidinvasion, Imbibitionfronts} to crack fronts in fracture mechanics \cite{CracklingDynamics, Crackpropagation}. Impurities and imperfections in realistic materials create interesting physics in these systems, as there is a competition between the local disorder, which makes the interface rough, and the elasticity, which tries to keep the interface smooth. For example, the fluctuating toughness of a material makes crack fronts irregular, as some parts of the crack advance before others. Neighboring segments try to keep up with each others' movements, and as a result, a weak point in the material can trigger movement in its whole neighborhood. These collective movements that reinforce themselves are called avalanches. \cite{FISHER1998113, StatisticalModelsofFracture}

The behaviour can be divided into three categories depending on the strength of the applied field or force. Under very small driving, the avalanches are initiated by thermal noise. This region of slow intermittent movement is called creep, and it is the region of interest in this paper. When the driving exceeds a critical value, called the depinning point, the movement becomes continuous, and the average velocity grows as a power of the driving. Lastly, under large external fields or forces the disorder becomes irrelevant, and the interface velocity grows linearly with the driving. A higher temperature blurs the transition points, causing for example depinning-like behaviour to start earlier. \cite{MiddletonThermalRounding, Bustingorry_2007, InterfaceMotioninRandom}

Avalanches and the average movement of interfaces have been studied a lot under constant driving. A varying driving is however interesting in for example magnetic hysteresis, where the applied magnetic field can alternate between opposite directions. Nattermann, Pokrovsky, and others derived theory for domain wall motion under periodic driving in a series of papers for depinning \cite{Nattermann1998, Nattermann2003} and for thermal motion \cite{Nattermann2001, Nattermann2004}. Recently the subject has gathered more attention, as a zero-average square wave magnetic field was found to change the roughness and size of circular magnetic domains \cite{Domenichini2019, Domenichini2021, Caballero_2021}.

In this paper, we calculate the average velocity of an interface under a sinusoidal external field in the creep region. We consider both when the elasticity is short-ranged, such as for magnetic domain walls, and long-ranged systems. Periodic driving in long-range elastic systems is directly relevant in the case of fatigue fracture, or fatigue crack propagation, as planar cracks are known in many cases to follow the long-range elastic interface model  \cite{Rice, GaoRice}. The full three-dimensional  situation is more complex, but the in-plane velocity of the front is expected to show similar dynamics \cite{10.3389/fphy.2014.00070}.
Fatigue can of course depend also on the presence and degree of plastic deformation, but our results should apply for very brittle materials and also help to connect fatigue and creep.

\section{Interface velocity}

We use the notation and terminology of magnetic domain walls in this section. The results still apply to elastic interfaces in other systems as well. In subsection \ref{LongRange} we however focus on long-range elastic interfaces and fracture mechanics and use corresponding terminology.

Several experiments have shown that thermally assisted domain wall velocity follows the creep behaviour for a large range of velocities until the driving field reaches its depinning value \cite{Lemerle, Jeudy2016,Jeudy2018,doi:10.1063/1.2337356, PhysRevLett.99.217208}. The creep velocity has been derived theoretically using energy arguments and an Arrhenius activation \cite{Nattermann1987, Ioffe1987} and later using the interface equation of motion and a functional renormalization group \cite{Chauve2000, Mueller2000}. The velocity under a magnetic field with strength $H$ is

\begin{equation}
    \label{eqn:creep}
    v_c = v_0 \exp({-CH^{-\mu}/T}),
\end{equation}
where $v_0$ and $C$ depend on the material and temperature and $\mu$ is a constant depending on the specific model. For magnetic domain walls $\mu\approx0.25$ \cite{Lemerle, Jeudy2016,Jeudy2018,doi:10.1063/1.2337356, PhysRevLett.99.217208, Chauve2000}.  The general form is
\begin{eqnarray}
    \label{eqn:myy}
    \mu = \frac{d-\alpha+2\zeta}{\alpha-\zeta},
\end{eqnarray}
where $d$ is the dimension of the interface, $\alpha$ is the range of the elasticity kernel, and $\zeta$ is the equilibrium roughness exponent \cite{Chauve2000}. Domain walls in thin systems are usually considered as one-dimensional. $CH^{-\mu}$ describes the energy scale needed to move the domain wall in a disordered landscape, and it diverges as the driving field vanishes.

Jeudy et al. proposed to express the creep velocity in terms of effective experimental variables for domain walls as 
\begin{equation}
    \label{eqn:Jeudy}
    v_c = v(H_d,T) \exp \Bigg(-\frac{T_d}{T} \Bigg[ \Bigg(\frac{H}{H_d} \Bigg)^{-\mu}-1 \Bigg] \Bigg),
\end{equation}
where $H_d$ is the field strength at which the domain wall velocity starts to resemble depinning instead of creep, $v(H_d,T)$ is the velocity at that field, $T$ is the temperature, and $T_d$ sets a material-dependent energy scale \cite{Jeudy2016, Jeudy2018}. $H_d$ and $v(H_d,T)$ also depend on the material and the temperature, as higher temperatures cause the depinning transition to start at smaller drivings. With this definition, the exponent
\begin{equation}
    \label{eqn:energybarrier}
    \frac{T_d}{T} \Bigg[ \Bigg(\frac{H}{H_d} \Bigg)^{-\mu}-1 \Bigg],
\end{equation}
vanishes at the effective depinning field $H_d$. Comparing Equations \ref{eqn:creep} and \ref{eqn:Jeudy} we see that \begin{equation}
    \label{eqn:v0vd}
    v_0 = v(H_d,T) e^{T_d/T}
\end{equation}
and
\begin{equation}
    \label{eqn:CTd}
    C=T_d H_d ^ \mu.
\end{equation}

The creep formula is all the theoretical background that we need for determining the interface velocity under a sinusoidal driving. We will mostly use the notation of Equation \ref{eqn:creep} in the following calculations, but for obtaining numerical results in subsection \ref{Numerical}, we use the notations of Jeudy et al. as we use material parameters from their publication \cite{Jeudy2018} to put the results in place in the context of domain walls.

\subsection{Sinusoidal driving}

If we replace the driving in the creep formula \ref{eqn:creep} with a periodic one $ H_0 + \Delta H \sin (\omega t)$, we can study the motion during a hysteresis loop, as is done in fatigue experiments for materials where the stress is varied in such a manner (often most sinusoidally or with a load protocol with alternating constant external stresses) integrating over the positive part of the cycle gives the expression

\begin{equation}
    \label{eqn:hysteresisintegral}
    v_p = \frac{v_0}{\pi} \int _{0} ^{\pi} \exp \Big(-\frac{C}{T} [H_0 + \Delta H \sin x]^{-\mu} \Big) dx
\end{equation}
for the mean velocity of the interface. Here $x=\omega t$. We consider whether the negative half-cycle from $\pi$ to $2\pi$ needs a separate treatment at the end of this subsection.

Using the methods shown in the Appendix, the integral can be approximated as 
\begin{equation}
    \label{eqn:finalvh}
    v_p \approx  \frac{2}{\pi a} e^b K_0(b) v_c(H_{max}),
\end{equation}
where $K_0$ is the modified Bessel function of the second kind, $v_c(H_{max})=e^{-CH_{max}^{-\mu}/T}$ is the creep velocity with field strength $H_{max}$, $a=([(3\mu+2)\Delta H-H_0]/H_{max})^{1/2}$, and $b= \mu C H_{max}^{-\mu} / [(3\mu+2-H_0/\Delta H)T]$.

The result can also be written using the series representation of $K_0$. Then
\begin{equation}
    \label{eqn:Bessel}
    v_p \approx  \Bigg(\frac{2H_{max}^{\mu+1}T}{\pi\mu C\Delta H}\Bigg)^{1/2} \Big( 1-\frac{1}{8b} + \frac{3^2}{2!(8b)^2} -\frac{3^2 5^2}{3!(8b)^3} +... \Big) v_c(H_{max}).
\end{equation}
The periodic velocity or the velocity under some other variable magnetic field can, of course, be used as an alternative way to measure the energy parameter $C$ without performing creep experiments with slower driving. If an experimentalist measures a creep velocity $v_c$ and a mean velocity $v_p$ under sinusoidal driving, the leading term in Equation \ref{eqn:finalvh} can be used to solve $C$ as
\begin{equation}
    \label{eqn:CLaplace}
    C = \frac{2 H_{max}^{\mu+1} T }{ \pi \mu \Delta H  } \frac{v_c^2(H_{max})}{v_p^2}.
\end{equation}
Although the right side of the expression includes the magnetic field, $C$ does not depend on it. Instead, $H_{max}^{\mu+1}/\Delta H$ cancels out the field dependence of $v_c$ and $v_p$.

What happens if the interface is driven for a whole cycle from $x=0$ to $x=2\pi$ instead of only for the positive half-cycle from $x=0$ to $x=\pi$? If $H_0 \geq \Delta H$, the driving remains positive for the whole cycle. First, we have to note that in the creep region the factor $C/T$ is large, as otherwise the interface movement would be fast and not creep. Because of the exponential form of the velocity and the large $C/T$ factor, the contribution of a negative half-cycle from $x=\pi$ to $x=2\pi$ is negligible compared to the contribution of the saddle point in the positive half-cycle. Indeed the result from our saddle-point approximation does not change. 

If $H_0<\Delta H$, the negative half-cycle includes a part with negative driving. In the case of a magnetic domain wall, negative driving means inverting the external magnetic field, which causes the interface to move backward. For fractures negative driving would mean compression, which is no longer a fracture line propagation problem. If $H_0$ is only slightly smaller than $\Delta H$, the saddle point in the positive half-cycle still dominates the overall movement. If on the other hand $H_0 = 0$, the positive and negative half-cycles cancel each other out, and the average velocity during both half-cycles is described by Equation \ref{eqn:finalvh}. This leaves us with the last option where there is significant and non-equal movement backwards and forwards. In that case, there is a saddle point for both half-cycles. Carefully going through the calculation in the Appendix shows that the final velocity is the difference between two velocities of the form of Equation \ref{eqn:finalvh}, where for the first one the maximal field is $H_{max}=\Delta H + H_0$, and for the second one $H_0$ is replaced by $-H_0$ in the expressions for $H_{max}$, $a$, and $b$.

\subsection{Circular magnetic domains}

Sometimes extra care is needed for alternating magnetic fields. Domenichini et al. found experimentally that circular magnetic domains reduce in size when an alternating field with zero average is applied. The curvature of the domain was found to induce an effective magnetic field inversely proportional to the radius of the magnetic bubble. \cite{Domenichini2019} A later computational study \cite{Caballero_2021} confirmed the result. This means that for a circular domain with radius $r$ an extra term proportional to $-1/r$ needs to be added to the magnetic field in our calculations. Assuming that the radius does not change much during one cycle, the extra term is a constant and can be included in $H_0$. The numerical study \cite{Caballero_2021} shows the evolution of the radius during the first few cycles, and indeed it changes only by a few percent between the minimum and maximum of a cycle, even though the simulated system is smaller and thus more sensitive than the ones used in experiments. 

If the duration of the cycles or the amplitude of the magnetic field is very large, but still within our creep framework, then of course the radius of a circular magnetic domain changes drastically during one cycle. In that case, the effective magnetic field induced by the curvature depends on how much the wall has already advanced during that cycle. Adding a term proportional to $-1/r=-1/(r_0+\int vdt^\prime)$ to the magnetic field $H_0 + \Delta H \sin x$ in Equation \ref{eqn:hysteresisintegral} complicates the integral significantly, so it might be easier to solve numerically.

\def\stackalignment{l}

\section{Experimental implications}

\subsection{Short-range elasticity}\label{Numerical}

\begin{figure*}

  \begin{subfigure}{.5\linewidth}
    \topinset{(a)}{\centering\includegraphics[width=\linewidth]{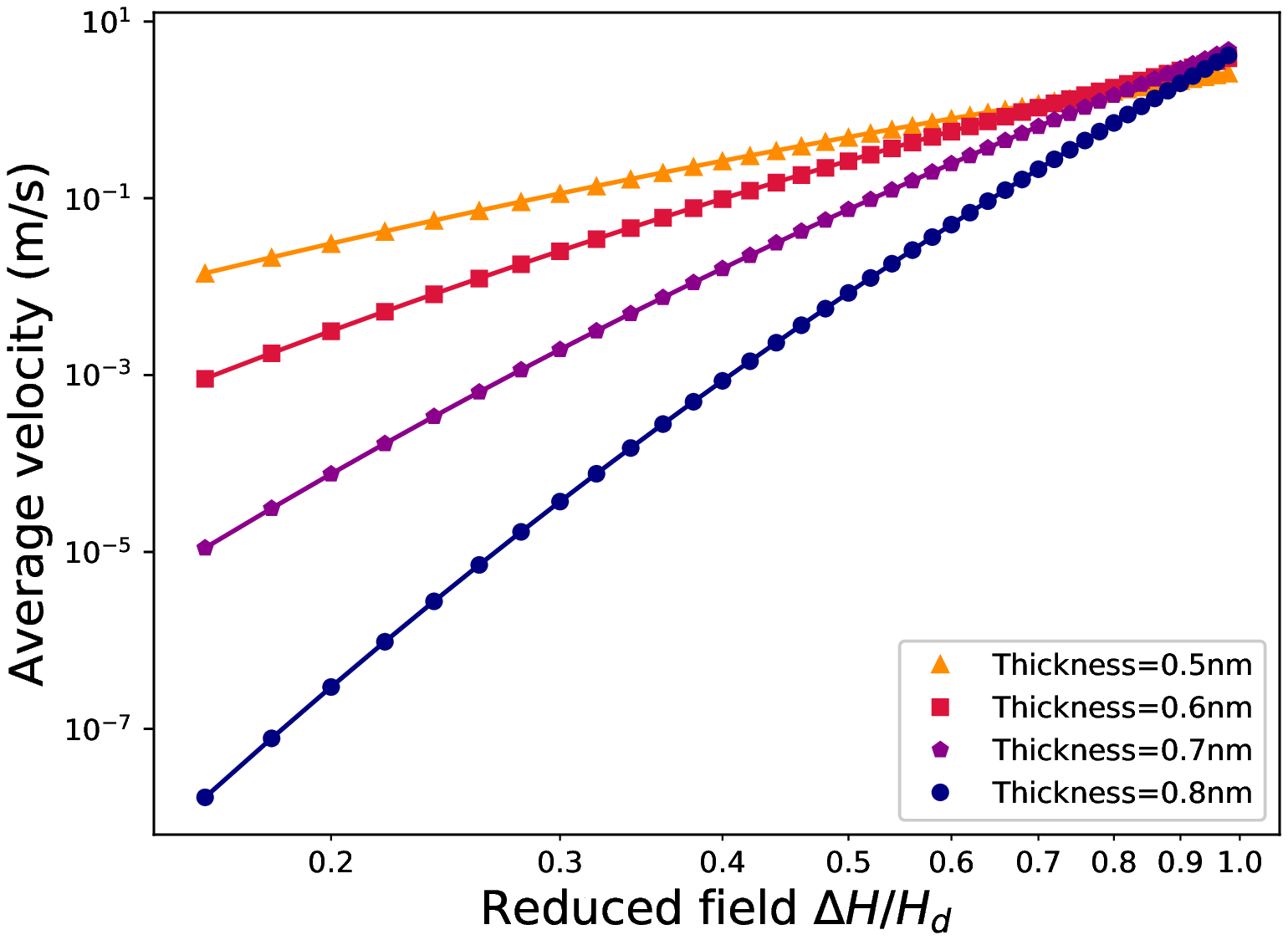}}{0.25in}{0in}
    \phantomcaption{\label{fig:velH00}}
  \end{subfigure}
  \begin{subfigure}{.5\linewidth}
    \topinset{(b)}{\centering\includegraphics[width=\linewidth]{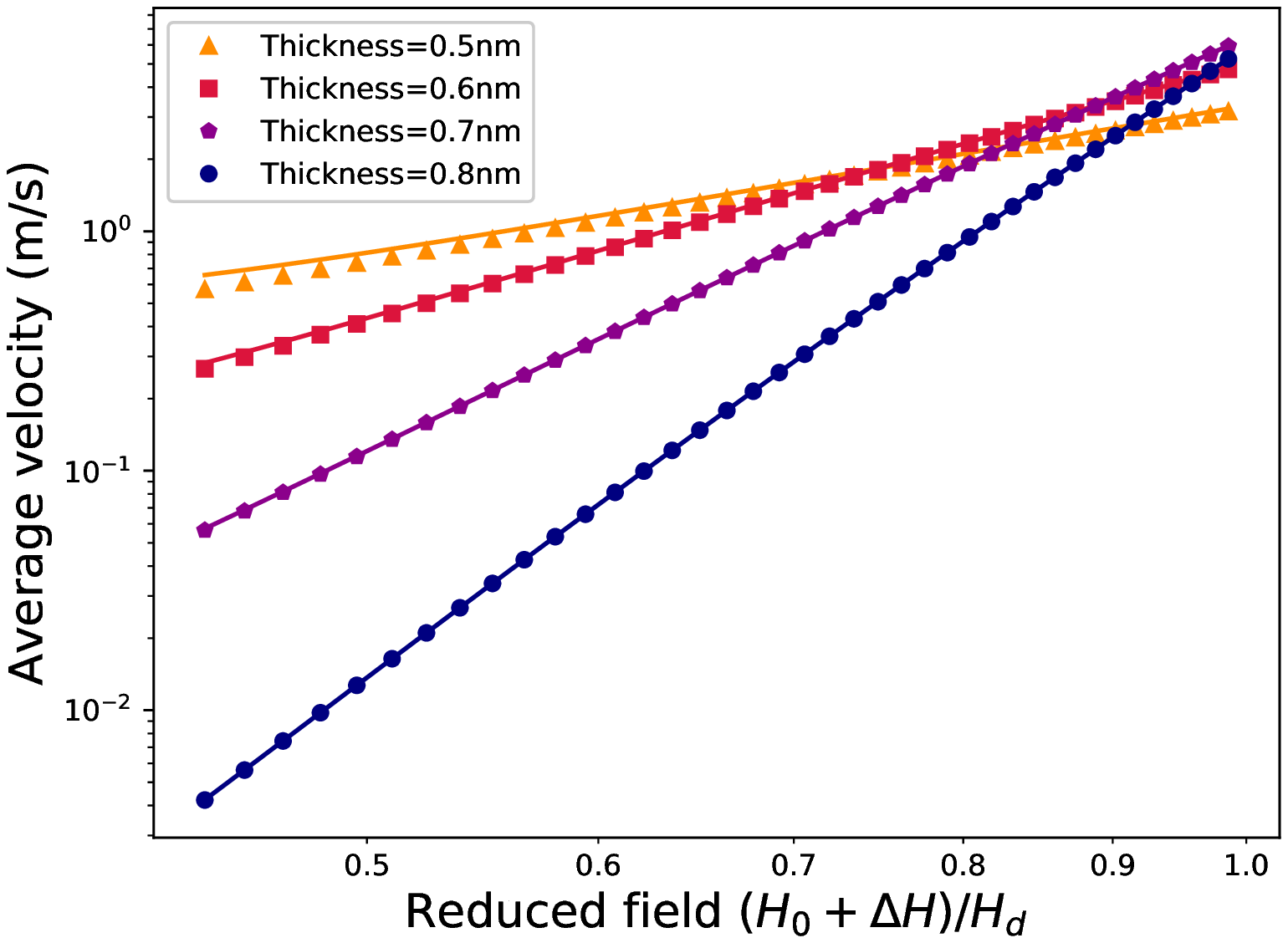}}{0.25in}{0in}
    \phantomcaption{\label{fig:velh0>0}}
  \end{subfigure}
  
  \caption{\label{fig:Velocity}The velocity of the domain wall with different minimum fields. On the left the minimum field is set to zero and on the right the minimum field $H_0=0.3H_d$. The temperature is $293K$ in both panels. The points denote the velocity obtained from a numerical integration of Equation \ref{eqn:hysteresisintegral} and the continuous lines use the approximation \ref{eqn:finalvh}. The data uses experimental values for Pt/Co/Pt films from \cite{Jeudy2018} and \cite{CreepandFlowRegimes}. The lines represent samples with different thicknesses. The values of $H_d$ are 28.5, 56, 76, and 72 in order of rising sample thickness. Notice that the axes are logarithmic.}

\end{figure*}

To check our formula for the domain wall velocity during a hysteresis loop, we integrated numerically expression \ref{eqn:hysteresisintegral} and compared it to the approximation \ref{eqn:finalvh} using material parameters for Pt/Co/Pt films from the table in \cite{Jeudy2018}, which uses experimental values from \cite{CreepandFlowRegimes}. Figure \ref{fig:Velocity} shows the domain wall velocity under various sinusoidal fields. The continuous lines use our approximation, and the points use the numerical integral. In Figure \ref{fig:velH00} the minimum field is $H_0=0$, whereas in Figure \ref{fig:velh0>0} $H_0=0.3H_d$, i.e., 30\% of the effective depinning value at room temperature. Different lines represent samples with different thicknesses.

\begin{figure*}

  \begin{subfigure}{.5\linewidth}
    \topinset{(a)}{\centering\includegraphics[width=\linewidth]{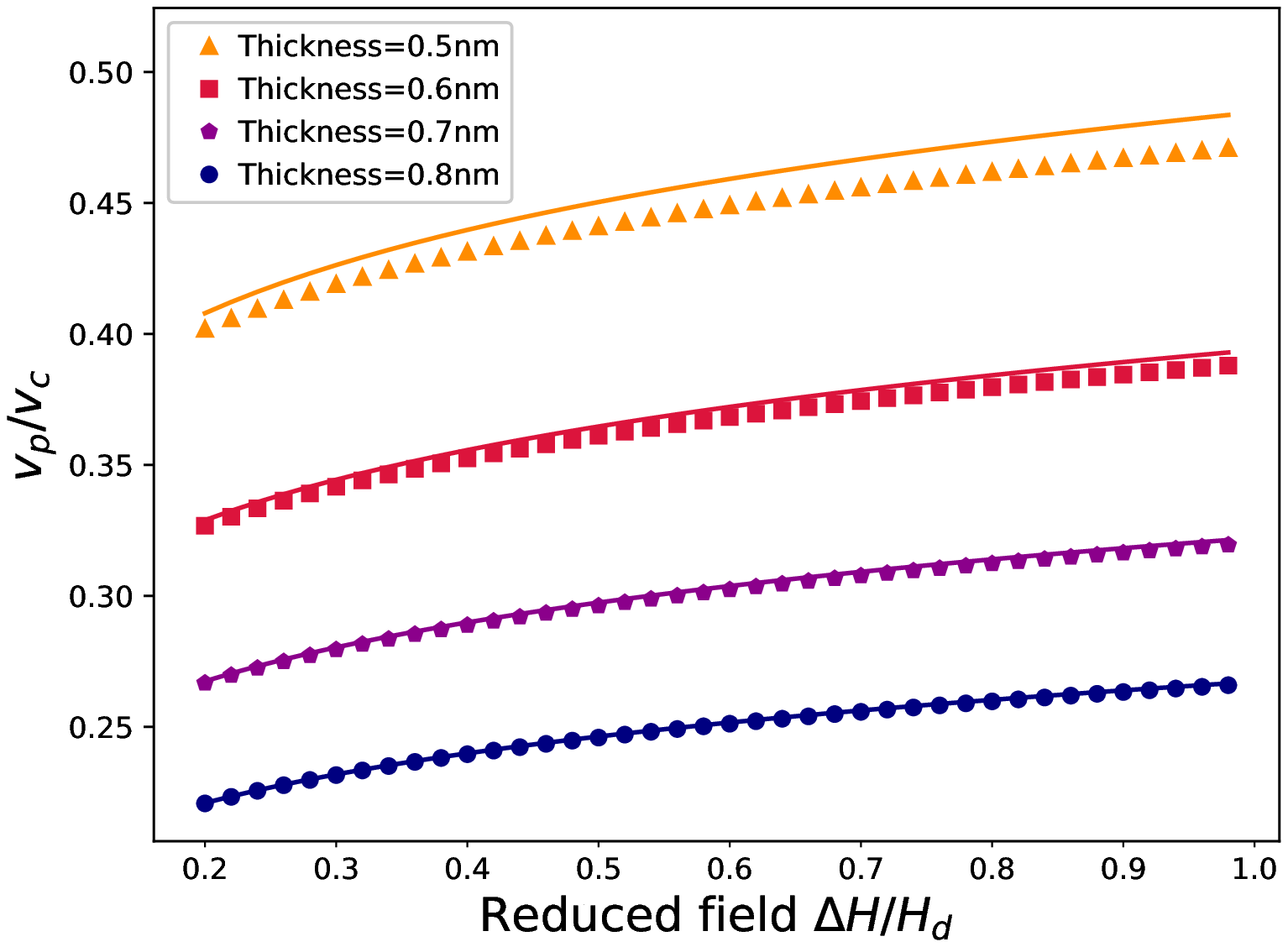}}{0.25in}{0in}
    \phantomcaption{\label{fig:BesselH00}}
  \end{subfigure}
  \begin{subfigure}{.5\linewidth}
    \topinset{(b)}{\centering\includegraphics[width=\linewidth]{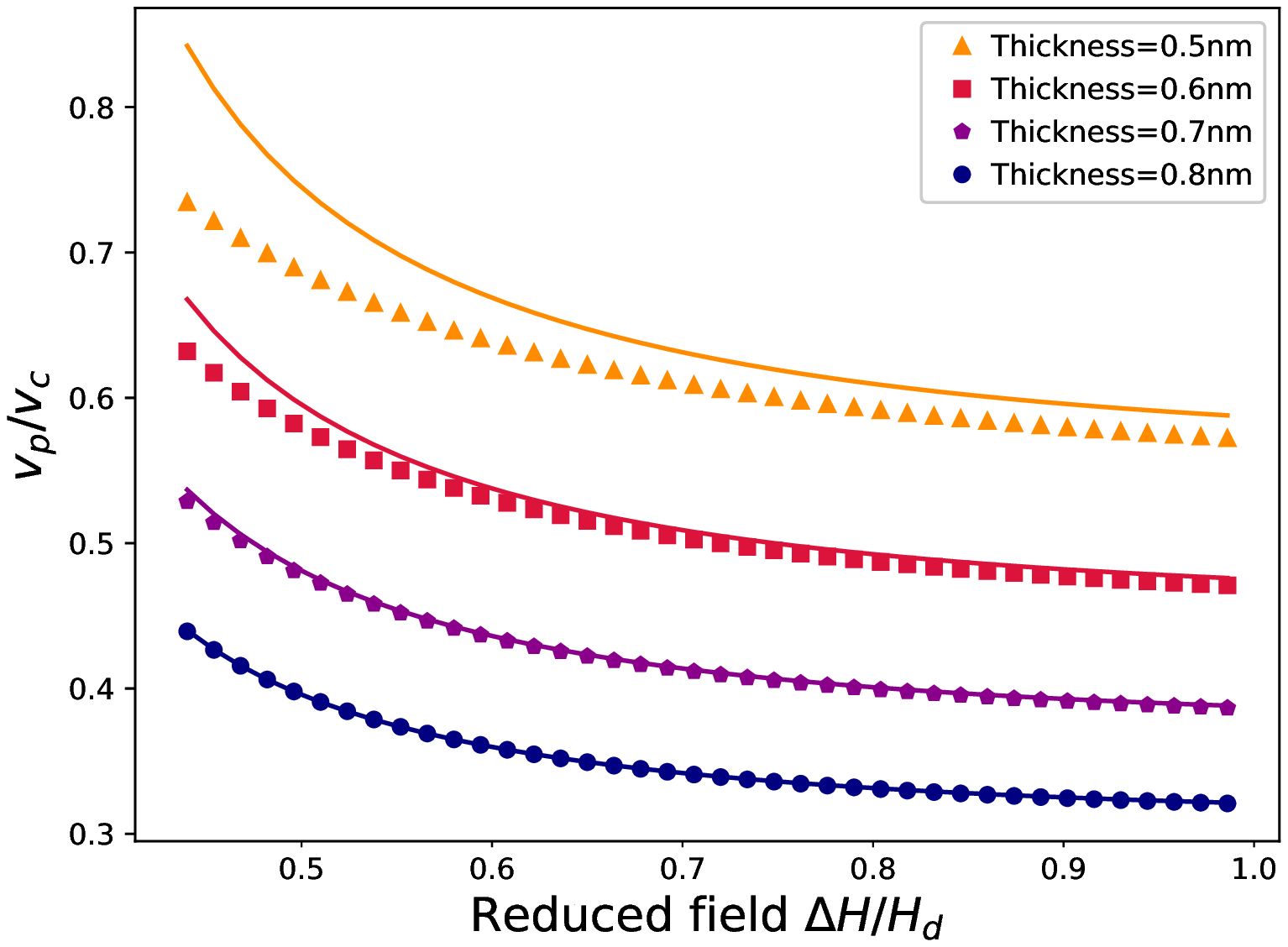}}{0.25in}{0in}
    \phantomcaption{\label{fig:BesselH0>0}}
  \end{subfigure}
  
  \caption{\label{fig:Bessel} The deviation of the periodic velocity from creep. The points use the numerically integrated velocities in Figure \ref{fig:Velocity} divided by the creep velocity $v_c(H_{max})=v_c(H_0+\Delta H)$ and the continuous lines use Equation \ref{eqn:finalvh}. In (a) the minimum field $H_0=0$ and in (b) $H_0=0.3H_d$. The error of Equation \ref{eqn:finalvh} for small fields in \ref{fig:BesselH0>0} results from using a saddle point approximation for a near constant field.}

\end{figure*}

Let us look at the behaviour of the function multiplying the creep velocity in Equation \ref{eqn:finalvh}. Figure \ref{fig:Bessel} shows the ratio of the periodic velocity to the creep velocity as a function of the maximal field $H_{max}$. The ratio is roughly constant, especially when the minimum field is zero, so the periodic velocity largely follows the creep formula \ref{eqn:creep}. The points use the previous numerical integration of Equation \ref{eqn:hysteresisintegral} and the continuous lines use Equation \ref{eqn:finalvh}. When the amplitude $\Delta H$ is smaller than the background field $H_0$, Equation \ref{eqn:finalvh} becomes less accurate. This is only natural, since for small amplitudes the field is closer to a constant, and we used a saddle point approximation.

\begin{figure*}

  \begin{subfigure}{.5\linewidth}
    \topinset{(a)}{\centering\includegraphics[width=\linewidth]{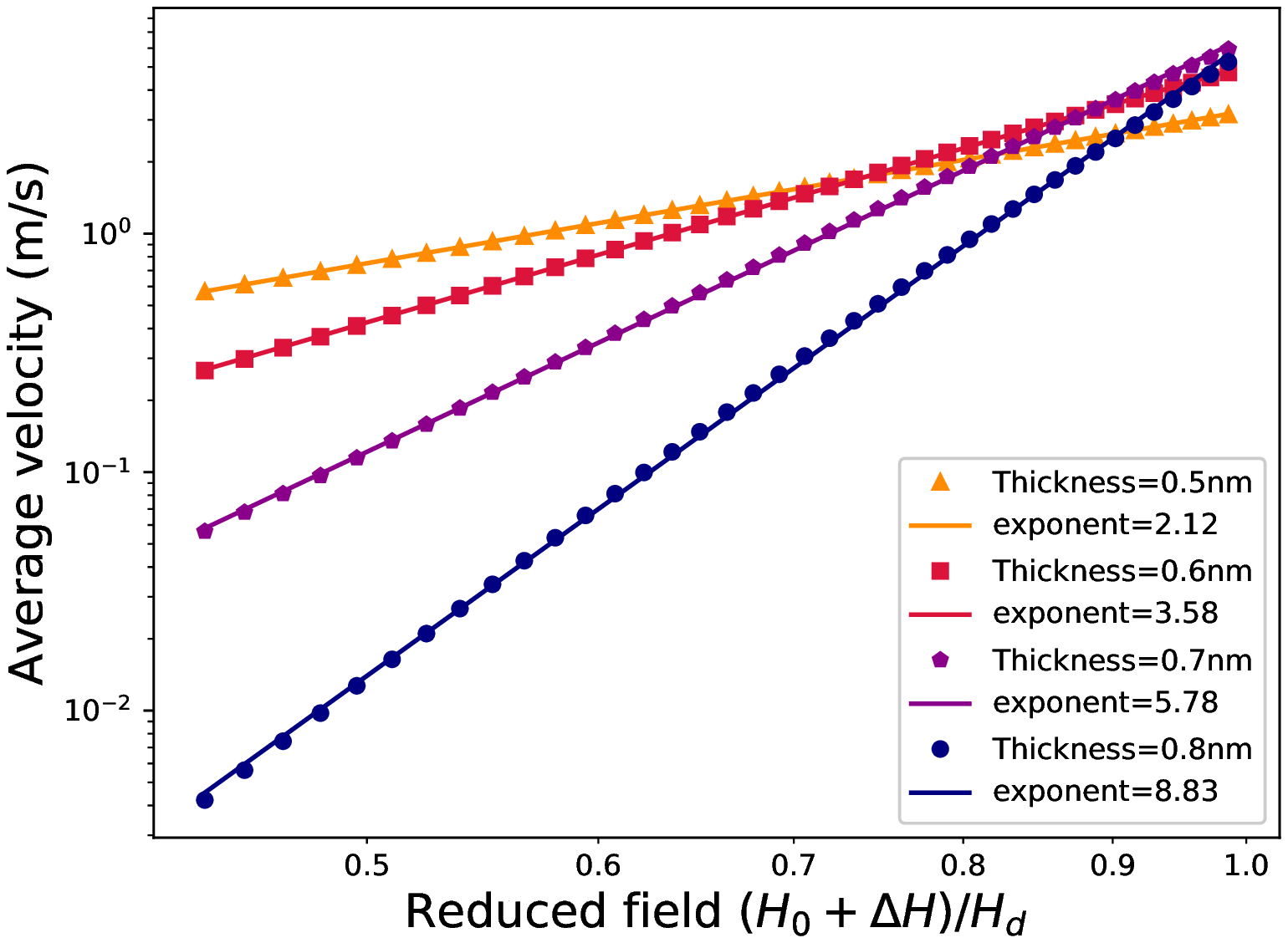}}{0.25in}{0in}
    \phantomcaption{\label{fig:plaw}}
  \end{subfigure}
  \begin{subfigure}{.5\linewidth}
    \topinset{(b)}{\centering\includegraphics[width=\linewidth]{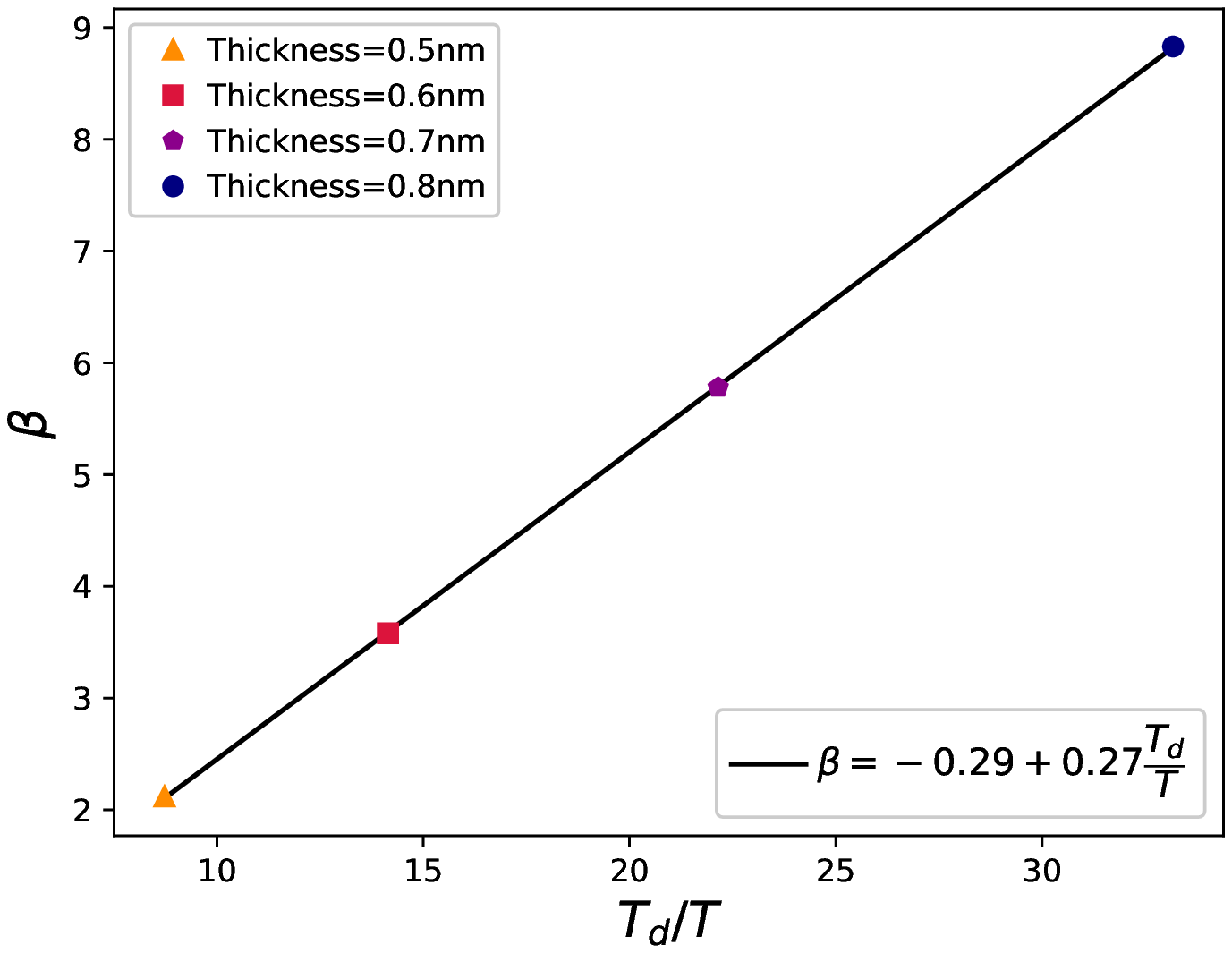}}{0.25in}{0in}
    \phantomcaption{\label{fig:betafit}}
  \end{subfigure}
  
  \caption{\label{fig:beta} (a) shows the numerically integrated velocity in Figure \ref{fig:velh0>0} fitted with a power law. The continuous lines are power-law fits using the function $\alpha x^\beta$. (b) shows the fitted exponents as a function of the material parameter $T_d=C/H_d^\mu$. The values of $T_d$ are 2558, 4145, 6490, and 9720 in order of rising sample thickness.}

\end{figure*}

The graphs for the periodic velocity, especially on Figure \ref{fig:velh0>0}, resemble very clean power laws, even though our approximation for the periodic velocity does not look like a power law. Indeed, Figure \ref{fig:plaw} shows the numerically integrated velocity of Figure \ref{fig:velh0>0} fitted with a simple power law $\alpha x^\beta$. Different material parameters lead to different exponents for the fit, as they change the energy scale $T_d$ and the effective depinning point $H_d$, resulting in a different value for $C$.

Since the periodic velocity is roughly proportional to the creep velocity in Figure \ref{fig:Bessel}, the apparent power-law behaviour arises mostly from the stretched exponential form of the creep velocity. However, periodic driving and the minimum field $H_0$ have an effect as well, as Figure $\ref{fig:velh0>0}$ with a positive minimum field resembles a cleaner power law than Figure \ref{fig:velH00} with $H_0=0$.

\newcommand{\appropto}{\mathrel{\vcenter{
  \offinterlineskip\halign{\hfil$##$\cr
    \propto\cr\noalign{\kern2pt}\sim\cr\noalign{\kern-2pt}}}}}

As $\mu\approx0.25$ for magnetic domain walls, we can do a simple estimate for the fitted exponents using the small $\mu$ approximation
\begin{eqnarray}
    \label{eqn:approximation}
    H^{-\mu} \approx 1- \mu \ln H.
\end{eqnarray}
Now the creep velocity $v_c=v_0 \exp(-CH^{-\mu}/T)$ turns it into the power law
\begin{eqnarray}
    \label{eqn:approximation2}
    v_c \propto H^{C\mu/T}.
\end{eqnarray}
Using $H_d$ and $T_d$ the relation is 
\begin{eqnarray}
    \label{eqn:approximation3}
    v_c \propto \Bigg( \frac{H}{H_d} \Bigg)^{\mu T_d / T}.
\end{eqnarray}
Indeed the fitted exponents in Figure \ref{fig:plaw} grow roughly as $-0.29+0.27T_d/T \approx \mu T_d /T$, as is illustrated in Figure \ref{fig:betafit}.

\subsection{Long-range elasticity}\label{LongRange}

Finally, we consider long-range elastic systems such as fractures, contact line wetting \cite{contactangle}, and low-angle grain boundaries \cite{grainboundaries}. For a one-dimensional long-range elastic interface $d=1$ and the elasticity kernel $\alpha=1$ in Equation \ref{eqn:myy}. Experiments where Plexiglas plates are attached by sintering and then detached mimicking a planar crack suggest that $\zeta \approx 0.35$ \cite{Santucci2008,Santucci2010}, which results in $\mu\approx1.08$. Similarly, a paper peeling experiment suggests that $\mu=1$ \cite{paperpaper}. Assuming that $\mu=1$, the creep law turns into a pure exponential form, and the leading term of the periodic velocity in Equation \ref{eqn:Bessel} becomes 
\begin{eqnarray}
    \label{eqn:longrange}
    v_p = \Bigg( \frac{2H_{max}^2 T }{ \pi C  \Delta H} \Bigg)^{1/2} v_0 e^{-C/H_{max}T}.
\end{eqnarray}
Now that $\mu=1$, we cannot do the approximation for small $\mu$, so the approximate power-law behavior with a material-dependent exponent is lost. Instead, the velocity increases proportionally to $\sqrt{H_{max}T/C} e^{-C/H_{max}T}$ when $\Delta H \gg H_0$.

\subsection{Fatigue cracks}
Fractures in brittle materials should follow
the behaviour in Equation \ref{eqn:longrange}. There is however a widely
known power-law relation between crack velocity and the
amplitude of an oscillating stress intensity factor known
as Paris’ law, where the exponent is not universal \cite{ParisErdogan,Ritchie}. The law states that the velocity is proportional to the Stress Intensity Factor (SIF) to some (empirical) power, and thus relates to the external stress (averaged over a cycle) with a power law. This contradicts our finding, which means that plastic deformation in the proximity of the crack line, which we ignored, must be more important for the periodically driven fatigue.
On the other hand, fractures in practical applications are
not planar, and the effective $\mu$-exponent may lead to qualitatively different behavior for the
crack velocity. In the case of an actually planar crack with no significant plastic deformations, such as in the Plexiglas experiments, our results should apply directly.

If fatigue were to follow the elastic interface model with
just a small value for the $\mu$-exponent, like in the short-range elastic case, then we could use the previous approximation, yielding again a power law with a material-dependent exponent. The exponent would
decrease with increasing temperature. This would be a testable hypothesis.
Interestingly, the research on how temperature affects the Paris’ law exponent and crack velocity is not unanimous. Some studies have found that a higher temperature indeed decreases
the exponent \cite{ELSHABASY2004305,Iost1993}, some found that a higher temperature at least affects the crack velocity \cite{MAKHLOUF1993163}, and some
found that temperature does not have a significant effect \cite{ArakereGoswamiKrohnRamachandran,Chan1987}. 

\section{Summary}

In conclusion, we studied the average velocity of an elastic interface under periodic subcritical driving and thermal noise. We considered both short-range elastic systems such as magnetic domain walls during a hysteresis loop and long-ranged systems such as fractures.

We started with the creep velocity of an elastic interface and derived an approximation for the resulting average velocity under a sinusoidal driving. The acquired velocity is a modification of the creep law. We also compared the velocity to a numerical integral, finding our approximation to be good for practical applications.

For short-range elastic systems, the periodic velocity grows approximately as a power of the driving field's strength. The behavior applies already for creep, as a creep velocity $\exp(-CH^{-\mu}/T)$ is roughly proportional to $H^{C\mu/T}$ for small $\mu$, but a cleaner power law emerges in the sinusoidal case, especially if there is a positive minimum field. As the exponent for the creep velocity depends on a material-specific energy scale, the exponent for the periodic velocity is also material dependent.

Our result for long-range elastic systems is simpler. Because of a different roughness exponent, the creep velocity is purely exponential, and as a result the periodic velocity increases as $\sqrt{\Delta H} e^{-1/\Delta H}$ with the amplitude $\Delta H$ of the driving field when the minimum field is small. Interestingly, planar cracks are known to follow the long-range elastic interface model, but fractures in general do not have a velocity of this form under periodic loading. Instead, periodically loaded cracks have power-law velocities with varying exponents. This resembles more our result for short-range elastic systems. This disagreement with our result might be caused by plastic deformation in periodic loading, but realistic fractures are also not necessarily planar and the real nature of the elastic interactions (due to screening) is not obvious. At the very least, our work should help in connecting studies of creep and the periodically loaded fatigue, so that it is easier to tell how much of the propagation of a fatigue crack can be attributed to brittle creep, and what is caused by some other mechanism. Our results should especially be helpful in experiments with very brittle materials since there is less plastic deformation, and in experiments that also connect creep and fatigue by using different loading amplitudes.

\ack
J.S and M.J.A. acknowledge support from the Academy of Finland (Center of Excellence program, 278367 and 317464) and M.J.A. from the European Union Horizon 2020 research and innovation programme under grant agreement No 857470 and from European Regional Development Fund via Foundation for Polish Science International Research Agenda PLUS programme grant No MAB PLUS/2018/8.
The authors acknowledge the computational resources provided by the Aalto University School of Science ``Science-IT'' project.

\appendix

\section{Derivation of the periodic velocity}

\begin{figure}
\includegraphics[width=\linewidth]{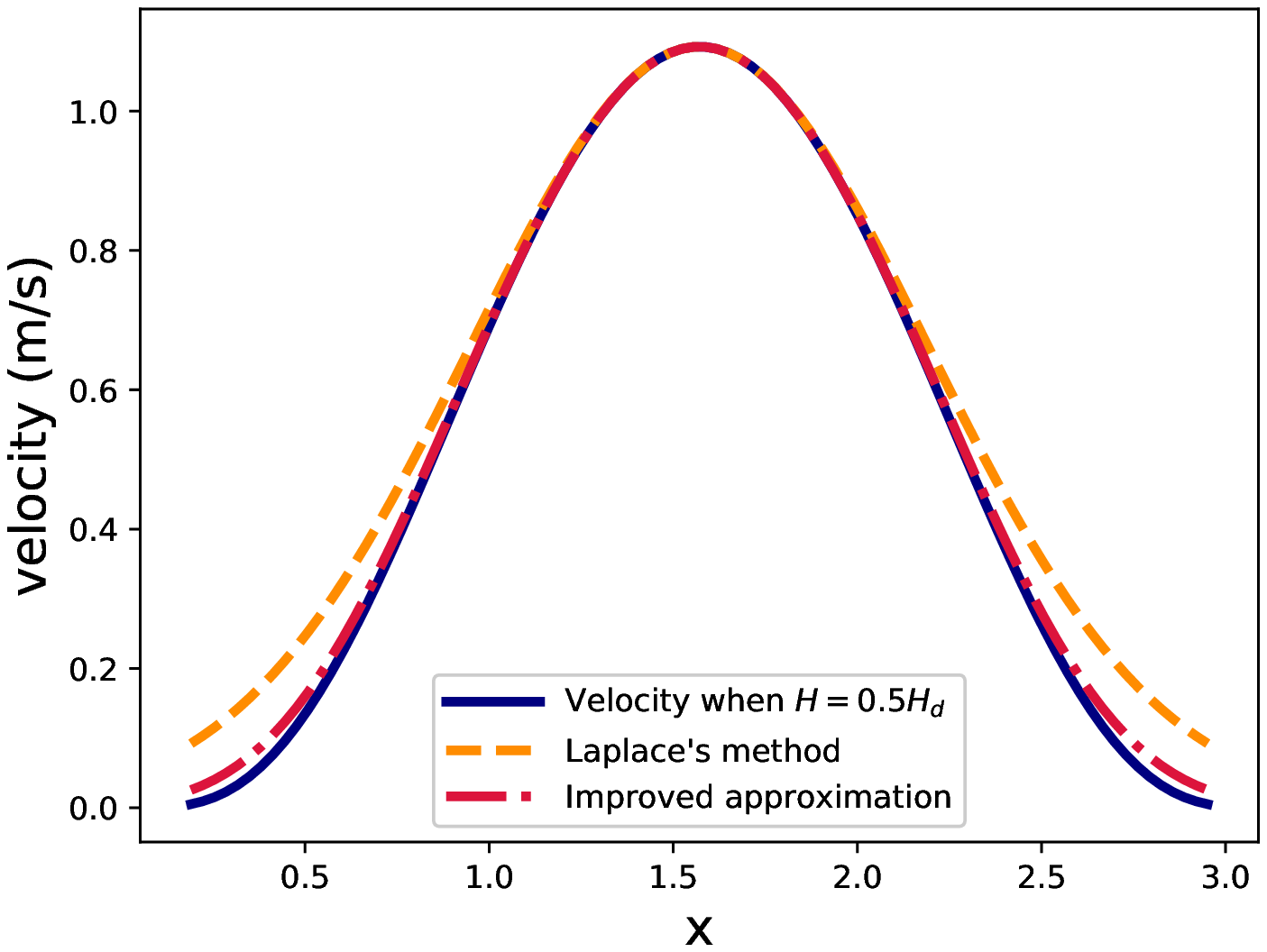}
\caption{\label{fig:approxvh} Illustration of the two approximations for the velocity $\exp(-C (\Delta H \sin x)^{-\mu}/T)$ during a cycle from $x=0$ to $x=\pi$. The values for $C$, $T$, and $H_d$ are from \cite{Jeudy2018} and \cite{CreepandFlowRegimes} for a Pt/Co/Pt film with thickness 0.5nm.}
\end{figure}

A simple way to approximate the integral in Equation \ref{eqn:hysteresisintegral} would be to use Laplace's method, which replaces the argument of the exponential function with an inverted parabola and the integration region with the whole real axis. We can improve the approximation by noticing that using the series representation of $C[H_0 + \Delta H \sin x]^{-\mu}/T$ around the saddle point at $x=\pi/2$, the exponential function in Equation \ref{eqn:hysteresisintegral} is of the form 
\begin{equation}
    \label{eqn:exponential}
	e^{-A-Bx^2-Dx^4-\mathcal{O}(x^6)},
\end{equation}
where the constants $A$, $B$, and $D$ are positive, and it can therefore be approximated with the function 
\begin{equation}
    \label{eqn:cosh}
    e^{-c-b\cosh(ax)}=e^{-c-b(1+a^2x^2/2+a^4x^4/24)-\mathcal{O}(x^6)},
\end{equation}
where the expansion is around $x=0$, by choosing $a=\sqrt{12D/B}$, $b=B^2/6D$, and $c=A-B^2/6D$ and modifying the integration boundaries. From the series representation of $C[H_0 + \Delta H \sin x]^{-\mu}/T$ at $x=\pi/2$ we see that $a=\big([(3\mu+2)\Delta H-H_0]/H_{max}\big)^{1/2}$, $b= \mu C H_{max}^{-\mu} / [(3\mu+2-H_0/\Delta H)T]$, and $c=CH_{max}^{-\mu} [1- \mu/(3\mu+2-H_0/\Delta H)]/T$.  There is however a divergence in $b$ and $c$, so $\Delta H$ has to be larger than $H_0/(3\mu+2)$. This is not a problem, since we already assume that the amplitude of the field $\Delta H$ is significantly larger than the background field $H_0$ by using a saddle-point expansion, and an amplitude of $H_0/(3\mu+2)$ or smaller would mean a near constant field. The method is illustrated in Figure \ref{fig:approxvh}.

Expanding both series further shows that the sixth order term in the series \ref{eqn:exponential} is positive, and in the series \ref{eqn:cosh} the term is also positive but smaller, so using a hyperbolic cosine is slightly better than just using a fourth order polynomial. In this case, a better approximation means that it can be used for smaller field amplitudes $\Delta H$.

Replacing the exponential function in integral \ref{eqn:hysteresisintegral} with the one in Equation \ref{eqn:cosh} and changing the integration region to the whole real axis as in Laplace's method, we get

\begin{eqnarray}
    \label{eqn:Aapproxintegral}
    \frac{v_c(H_{max})}{\pi} e^b \int _{-\infty} ^{\infty} \exp (b \cosh(a x) ) dx,
\end{eqnarray}
where $v_c(H_{max})=\exp(-CH_{max}^{-\mu}/T)$ is the creep velocity with field strength $H_{max}$.
This integral is one of the definitions of the modified Bessel function of the second kind $K_0$. Therefore, we acquire the average velocity
\begin{equation}
    \label{eqn:Afinalvh}
    v_p \approx  \frac{2}{\pi a} e^b K_0(b) v_c(H_{max})
\end{equation}
for an interface under periodic driving.

If we use the series representation of $K_0$ and write $a$ in terms of $b$, the result in Equation \ref{eqn:Afinalvh} takes the form
\begin{equation}
    \label{eqn:ABessel}
    v_p \approx  \Bigg(\frac{2H_{max}^{\mu+1}T}{\pi\mu C\Delta H} \Bigg)^{1/2} \Big( 1-\frac{1}{8b} + \frac{3^2}{2!(8b)^2} -\frac{3^2 5^2}{3!(8b)^3} +... \Big) v_c(H_{max}).
\end{equation}
The first term is the result that we would get using Laplace's method.

\section*{References}
\bibliography{iopart-num}

\end{document}